\documentclass[preprints,article,accept,moreauthors,pdftex]{Definitions/mdpi} 

\firstpage{1} 
\pubvolume{1}
\issuenum{1}
\articlenumber{0}
\pubyear{2021}
\copyrightyear{2020}
\datereceived{} 
\dateaccepted{} 
\datepublished{} 
\hreflink{https://doi.org/} 

\Title{Shaping Dynamical Casimir Photons}

\TitleCitation{Title}

\Author{Diego A. R. Dalvit $^{1}$*
and Wilton J. M. Kort-Kamp $^{2}$}

\AuthorNames{Diego A. R. Dalvit and Wilton J. M. Kort-Kamp}

\AuthorCitation{Diego A. R. Dalvit and Wilton J. M. Kort-Kamp}

\address{%
$^{1}$ \quad Los Alamos National Laboratory, MS B213, Los Alamos, NM 87545, USA; dalvit@lanl.gov\\
$^{2}$ \quad  Los Alamos National Laboratory, MS B262, Los Alamos, NM 87545, USA; kortkamp@lanl.gov}

\corres{Correspondence: dalvit@lanl.gov}

\abstract{Temporal modulation of the quantum vacuum through fast motion of a neutral body or fast changes of its optical properties is known to promote virtual into real photons, the so-called dynamical Casimir effect. 
Empowering modulation protocols with spatial control could enable to shape the spectral, spatial, spin, and entanglement properties of the emitted photon pairs. Space-time quantum metasurfaces have been proposed as a platform to realize this physics via modulation of their optical properties. Here, we report the mechanical analog of this phenomenon by considering systems whose lattice structure undergoes modulation in space and in time.
We develop a microscopic theory that applies both to moving mirrors with modulated surface profile and  atomic array meta-mirrors with perturbed lattice configuration. Spatio-temporal modulation enables  
motion-induced generation of steered frequency-path entangled photon pairs in co- and cross-polarized states, as well as vortex photon pairs featuring frequency-angular momentum entanglement.
The proposed space-time dynamical Casimir effect can be interpreted as an induced dynamical asymmetry in the quantum vacuum.}


\begin{document}

\section{Introduction}
The generation of photon pairs out of the quantum vacuum, the so-called dynamical Casimir effect (DCE),
was originally described as a motion-induced phenomenon \cite{Moore1970}, but it can occur when any kind of temporal modulation is exerted on the vacuum to promote virtual into real photons
\cite{Dodonov2010,Dalvit2011,Nation2012,Dodonov2020}.  Motion-induced photon generation has not been observed to date because it requires unfeasibly large modulation frequencies of a mechanical boundary, and several analog DCE systems have been demonstrated involving temporal modulation of material properties
\cite{Wilson2011,Jaskula2012,Lahteenmaki2013,Vezzoli2019}. Still, the
physics of motional dynamical Casimir effects offers interesting insights into the interplay between matter and field fluctuations in non-equilibrium systems. Motional DCE (also known as motion-induced  or mechanical DCE) is typically described in a ``field-centric" approach based on quantum fluctuations of the electromagnetic field supplemented with time-dependent boundary conditions. A ``matter-centric" approach has been recently pursued based on microscopic models that emulate a moving dielectric mirror as a collection of accelerated dipoles that emit quantum radiation \cite{Souza2018}. Interestingly, the two descriptions result in identical predictions for the angular emission profile of DCE photons. This duality between field-centric and matter-centric approaches also occurs in equilibrium fluctuation-induced interactions \cite{Milonni1994}.  Microscopic models can be used to study the dissipative counterpart of DCE emission, namely the drag force on the moving mirror, as well as the related problem of quantum friction and associated near-field DCE emission of surface polaritons \cite{Barton2010, Intravaia2015}. 

Modulation protocols having both temporal and spatial control can enable novel functionalities such as the generation of complex structured Casimir light. We have recently proposed an analog dynamical Casimir effect  based on the concept of space-time quantum metasurfaces \cite{STQM}, in which the optical properties of a quantum metasurface are modulated in space and time and generate DCE photon pairs with tailored spatial profiles at giant production rates.  Here, we develop the motion-induced version of this effect based on a microscopic model of a spatio-temporally modulated mirror. We describe the mirror as a collection of dipoles whose center-of-mass coordinates are modulated in space and in time, and find that photon pairs are generated out of vacuum with tailored spatial modes determined by the spatial modulation protocol. The collection of dipoles behaves as a quantum phase array antenna that emits structured Casimir light. The same microscopic theory applies to a quantum meta-mirror comprising an atomic array with modulated lattice structure  \cite{Bekenstein2020}. We call the proposed effect space-time DCE to distinguish it from the standard DCE process that involves temporal modulation only.
%
\section{Materials and Methods}

\subsection{Microscopic model for space-time motional DCE}

A microscopic model for a spatio-temporally modulated dielectric mirror consists of an array of $N$ multi-level atoms, each of which undergoes its own accelerated trajectory ${\bf R}_j(t)$ and couples to the electromagnetic field via the dipolar interaction. The trajectories are driven by some external agent and in this work we consider oscillatory trajectories typically employed in studies of the dynamical Casimir effect. We note, however, that our theory can be extended to arbitrary trajectories. The spatial modulation is incorporated via a 
synthetic phase $\Phi({\bf R}_{j})$ that is imprinted by a temporal delay of the oscillation of each atom. 
We thus consider
\begin{equation}
{\bf R}_j(t)={\bf R}_{j}+ \hat{\bf z}  \Delta \cos(\Omega t- \Phi({\bf R}_{j})), 
\label{motion}
\end{equation}  
where ${\bf R}_{j}$ is the static position of the $j$-th atom ($j=1, ..., N)$, 
$\Delta$ is the oscillation amplitude, and $\Omega$ the oscillation frequency.
A conceptual representation of the spatio-temporally modulated atoms is shown in Fig. 1. 
For example, a traveling-wave modulation on the $x-y$ plane 
corresponds to $\Phi({\bf R}_{j})=\boldsymbol{\beta} \cdot {\bf r}_{j}$ 
with ${\bf r}_{j}=(x_{j},y_{j})$ and $\boldsymbol{\beta}=(\beta_x,\beta_y)$ a  momentum ``kick" on the same plane. A spinning-wave modulation around an axis orthogonal to the mirror's plane corresponds to $\Phi({\bf R}_{j})=\ell \varphi_{j}$, where $\ell$ is an integer denoting an imprinted angular momentum and 
$\varphi_{j}$ is the azimuthal coordinate of each atom. Note that under this modulation the atoms are not rotating, just the temporal dephasing of their oscillatory motion along the $z$ direction is spinning around the $z-$ axis. The traveling-wave modulation could be implemented by an acoustic perturbation that launches a plane-wave on the surface of the mirror. In the case of a meta-mirror formed by an atomic array monolayer trapped in a  three-dimensional deep optical lattice, the modulation could be accomplished by shaking the vertical lattice and using spatial light modulators to set a phase shift among the vertical trapping fields on different atoms to imprint the linear synthetic phase.

\begin{figure}[t!]
\includegraphics[width=0.9\linewidth]{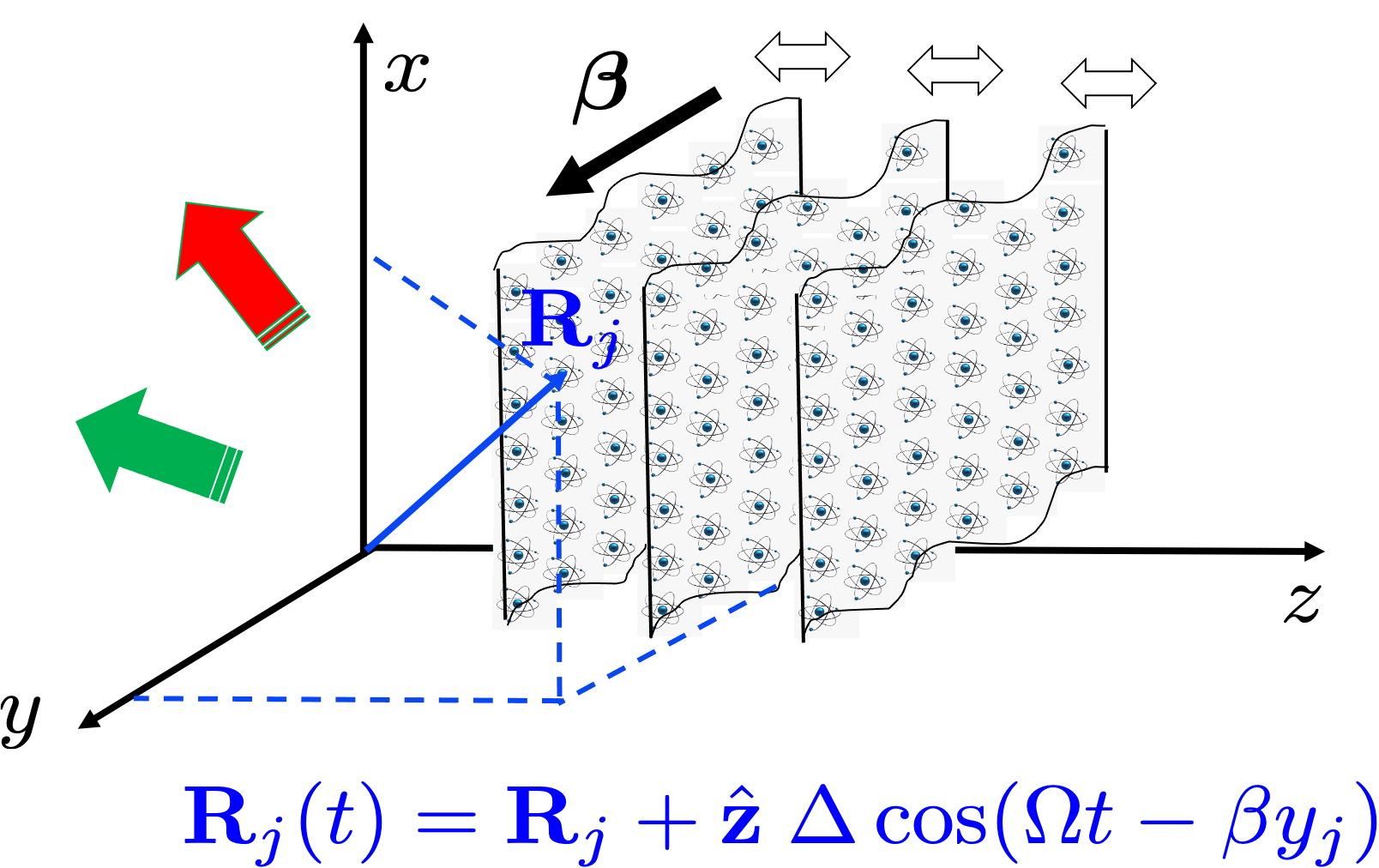}
\caption{
{\bf Concept of space-time motion-induced DCE.} 
An atomic array  is externally driven by a spatio-temporal modulation of their center-of-mass coordinates. Atoms oscillate along the $z$-direction and are temporally dephased by a linear synthetic phase 
distribution $\Phi({\bf R})=\beta y$. The modulation produces a traveling wave of ripples moving along the $y$-direction. Pairs of photons (red and green arrows) are emitted with in-plane linear momentum adding up to the momentum kick $\beta$.
}
\label{Fig1}
\end{figure}

The dipole array interacts with the electromagnetic field via the Hamiltonian (SI units used throughout)
\begin{equation}
H(t) \!=\!  - \sum_j {\bf d}_j(t) \cdot \!\Big[ {\bf E}({\bf R}_j(t),t) + {\bf v}_j(t)\times  {\bf B}({\bf R}_j(t),t) \Big], 
\end{equation}
where ${\bf d}_j(t)$ is the electric dipole operator of atom $j$, 
${\bf E}({\bf R},t)$ and ${\bf B}({\bf R},t)$ are the electric and magnetic fields both evaluated at the atomic positions, 
and ${\bf v}_j(t) =- \hat{\bf z} \Omega \Delta \sin(\Omega t- \Phi({\bf R}_j))$ is the (non-relativistic) 
velocity of each atom. In this paper the atomic internal degrees of freedom and the electromagnetic field are treated as quantum dynamical variables, while the atomic center-of-mass is a classical prescribed motion. 
The first term in the Hamiltonian is the usual electric dipole interaction and the second is the so-called R\"ontgen interaction. We note that the R\"ontgen term must be taken into account in the DCE far-field emission of photons, as we study here. For the DCE problem of emission in the near-field, e.g., pairs of surface polaritons induced on a surface by a close-by moving atom, it can be neglected \cite{Barton2010, Intravaia2015}. The non-zero matrix elements of the electric dipole operator ${\bf d}_j(t)$ in the interaction picture are
\begin{equation}
\langle g_j | {\bf d}_j(t)| e_j \rangle = \hat{\boldsymbol{\eta}}_j d_{eg} e^{-i \omega_{eg} t} ,
\end{equation}
where $|g_j \rangle$ is the ground state of atom $j$, 
$|e_j \rangle$ is an excited state, $\hat{\boldsymbol{\eta}}$ is a real unit vector
 denoting the orientation of the $j$-th dipole, and $d_{eg}$ is the matrix element. Note that all atoms have the same $d_{eg}$ as we assume identical atoms. When the atoms are isotropic, the sum over orientations of any given atom gives $\sum_{\hat{\boldsymbol{\eta}}_j} (\eta_j)_a (\eta_j)_b = \delta_{ab}$. We consider that the atoms are sufficiently spaced within the array to neglect photon multi-scattering among different atoms, rendering the evolution of different atoms identical irrespective of their location within the array. 
The electromagnetic field in the interaction picture is then simply given by the free field, that we expand in a set of modes in the usual way
\begin{eqnarray}
{\bf E}({\bf R},t) &=& i \sum_{\gamma} \Big(\frac{\hbar \omega_{\gamma}}{2\epsilon_0 V}\Big)^{1/2} \; 
[ {\bf E}_{\gamma}({\bf R}) a_{\gamma} e^{-i \omega_{\gamma} t} - {\bf E}^*_{\gamma}({\bf R}) a^{\dagger}_{\gamma} e^{i \omega_{\gamma} t } ], \cr
{\bf B}({\bf R},t) &=& \sum_{\gamma} \Big(\frac{\hbar }{2\epsilon_0 V \omega_{\gamma} }\Big)^{1/2} \;  
[ \nabla \times {\bf E}_{\gamma}({\bf R}) a_{\gamma} e^{-i \omega_{\gamma} t} + 
 \nabla \times {\bf E}^*_{\gamma}({\bf R}) a^{\dagger}_{\gamma} e^{i \omega_{\gamma} t} ]  .
\end{eqnarray}
Here, $V$ is a quantization volume, 
${\bf E}_{\gamma}({\bf R})$ is the spatial mode,
$\omega_{\gamma}>0$ is the mode frequency, and
$a^{\dagger}_{\gamma}$, $a_{\gamma}$ are creation and annihilation operators of photons in mode $\gamma$.
The specific choice of modes is determined by the symmetries of the synthetic phase, as we discuss later in the paper.  

\subsection{Two-photon emission rate}

We assume that the initial state of the $N$-atom system plus electromagnetic field is all the (identical) atoms in their ground state and the field in vacuum, i.e.,
$|\psi(0) \rangle = | \{g \}; vac\rangle$, where we denote $| \{g \} \rangle = | g_1, g_2, ..., g_N\rangle$ the multi-atom ground state. The time-evolved state in the interaction picture to second order
in the dipolar couplings $d_{eg}$ is
\begin{eqnarray}
&& |\psi(t) \rangle \approx c_0(t) | \{g\} ;vac\rangle +  \sum_{j} 
 \sum_{e_j, \hat{\boldsymbol{\eta}}_j}
 \sum_{\gamma}  c^{(e_j)}_{\gamma}(t) 
|e_j,  \{g\}_j  ; \gamma \rangle \cr
&&+ \frac{1}{2} \sum_{j \neq j'} \sum_{e_j, \hat{\boldsymbol{\eta}}_j} \sum_{e_{j'}, \hat{\boldsymbol{\eta}}_{j'}}
\sum_{\gamma_1,\gamma_2} c^{(e_j,e_{j'})}_{\gamma_1,\gamma_2}(t) |e_j, e_{j'}, \{g\}_{j,j'}; \gamma_1,\gamma_2\rangle \cr
&&
+ \frac{1}{2} \sum_{j} \sum_{e_j, \hat{\boldsymbol{\eta}}_j} 
\sum_{\gamma_1,\gamma_2} c^{(e_j)}_{\gamma_1,\gamma_2}(t) |e_j, \{g\}_{j}; \gamma_1,\gamma_2\rangle 
+
 \frac{1}{2}
\sum_{\gamma_1,\gamma_2} c_{\gamma_1,\gamma_2}(t) | \{g\}; \gamma_1,\gamma_2\rangle.
\end{eqnarray}
The first term is a correction to the amplitude of the initial state due to the coupling. 
The second term corresponds to one photon emitted and the atomic array being in an equal superposition of one atom excited in level $e_j$ and all the rest in the ground state.
The third, fourth, and fifth terms are two-photon contributions in which two, one and no atom is excited, respectively.
We denote $|e_j, \{g \}_j \rangle = |g_1, ..., g_{j-1}, e_j, g_{j+1},... , g_N \rangle$ the state in which one atom is excited, and
$|e_j, e_{j'}, \{g \}_{j,j'} \rangle = |g_1, ..., g_{j-1}, e_j, g_{j+1},... , g_{j'-1}, e_{j'}, g_{j'+1}, ..., g_N \rangle$ the state in which two atoms are excited.
We make the {\it crucial assumption} that the modulation frequency is much smaller than any atomic transition
frequency from the ground state, $\Omega \ll \omega_{eg}$. In this approximation, the second, third, and fourth terms give transition amplitudes that are non-resonant and their contributions to the process of one- and two-photon generation is negligible and are discarded in the following. In this regime, the DCE process is solely given by the fifth term, for which photon pairs are generated and
 all atoms remain in their ground state. 

One can compute the transition amplitude $c_{\gamma_1,\gamma_2}(t)$ in two ways: 
(a) use second-order time-dependent perturbation theory based on the  bilinear Hamiltonian $H(t)$ that depends on field and atomic degrees of freedom, or 
(b) use first-order time-dependent perturbation theory based on an effective Hamiltonian $H_{eff}(t)$ that depends quadratically on field degrees of freedom and contains the information about the atoms through their ground-state polarizability. Approach (b) was used in \cite{Souza2018} for the case of a single atom and no synthetic phase: the atom-field interaction is written in the atom's comoving frame as the standard static dipolar interaction, it is then re-written as an effective Hamiltonian that traces over the atom's internal degrees of freedom rendering the interaction quadratic in the field and depending on the atomic ground-state dynamic polarizability, and finally the Hamiltonian is Lorentz boosted to the lab frame to get $H_{eff}(t)$. Although approach (b) is
technically simple,
it has the drawback that it obscures the joint atom-field dynamics. Furthermore, for multiple atoms and non-zero synthetic phases it requires to introduce multiple comoving frames because atoms in the array have different instantaneous velocities. 
For these reasons, in this work we prefer to follow the physically more transparent albeit more cumbersome approach (a). 

The transition amplitude is given by the usual expression in second-order time-dependent perturbation theory
\begin{equation}
c_{\gamma_1,\gamma_2}(t) \! =\! - \frac{1}{\hbar^2} \!\int_0^t \!\!dt' \! \!\int_0^{t'} \!\!dt'' \!
\sum_{j} \sum_{e_j, \hat{\boldsymbol{\eta}}_j \gamma}  \!\!
\langle \{g\}; \gamma_1,\gamma_2 | H(t') | e_j,  \{ g \}_j  ; \gamma \rangle
 \langle  e_j,  \{g\}_j  ; \gamma  | H(t'') | \{g\}; vac\rangle
\label{second}
\end{equation}
that sums over intermediate virtual states $| e_j,  \{ g \}_j  ; \gamma \rangle$.
In the limit when the relevant field wavelengths are much larger than the atomic displacements,
$\omega\Delta/c\ll 1$, we can approximate the R\"ontgen term of the Hamiltonian evaluating the fields at the static atomic positions, so that it has a time-dependency  $\sim \sin(\Omega t -\Phi({\bf R}_j))$ arising from the velocity.  Also, the dipolar term in the Hamiltonian can be expanded around the equilibrium positions, giving a time-dependency 
$\sim \cos(\Omega t - \Phi({\bf R}_j))$ from the spatial gradient of the field.  
After these approximations, the concatenated time integrals in Eq. (\ref{second}) can be readily evaluated  as they are reduced to integrals of products of simple harmonic functions. The final result is a sum of various terms that oscillate in time. We drop the co-rotating term $e^{i (\omega_1+\omega_2+\Omega)t}$ because it averages out to zero. Counter-rotating terms of the form $(\pm \omega_{1,2} \pm \Omega \mp \omega_{eg})^{-1} \times e^{i (\pm \omega_{1,2} \pm \Omega \mp \omega_{eg})t}$ and $(\pm \omega_{1,2} \pm \omega_{eg})^{-1} \times e^{i (\pm \omega_{1,2} \pm \omega_{eg})t}$ could potentially produce a resonant effect if their denominator vanished. However, under the assumption that the modulation frequency is much smaller than the atomic transition frequencies from the ground state, 
$\Omega \ll \omega_{eg}$, this cannot occur
(note that DCE photons necessarily have frequencies  $\omega_{1,2} \le \Omega$ per energy conservation). The only counter-rotating term contained in the transition amplitude that can produce a resonant effect is 
$(\omega_1+\omega_2-\Omega)^{-1} \times e^{i (\omega_1+\omega_2-\Omega)t}$. Upon using all these considerations, the transition amplitude results
\begin{eqnarray}
c^{(\Phi)}_{\gamma_1,\gamma_2}(t) &\approx & e^{i (\omega_1+\omega_2-\Omega)t/2}
  \frac{\Omega \Delta \sqrt{\omega_1 \omega_2}}{2c V} 
\Big( \sum_{e} \frac{2 d_{eg}^2}{3 \epsilon_0 \hbar \omega_{eg}} \Big)  \frac{\sin[ (\omega_1+\omega_2-\Omega)t/2]}{\omega_1+\omega_2-\Omega}
\cr
&& \times
\sum_j e^{i \Phi({\bf R}_j)}  W^*_{\gamma_1,\gamma_2}({\bf R}_j) ,
\end{eqnarray}
where we already performed the sum over orientations of the isotropic dipoles. Here,
\begin{eqnarray}
W_{\gamma_1,\gamma_2}({\bf R}_j)=\! \frac{c}{\Omega}  \hat{{\bf z}}\! \cdot\! 
\nabla_j [{\bf E}_{\gamma_1}({\bf R}_j)\! \cdot\! {\bf E}_{\gamma_2}({\bf R}_j)] 
-   \hat{\bf z}\! \cdot\!  [ \frac{c}{\omega_2} {\bf E}_{\gamma_1}({\bf R}_j) \!\times \! 
( \nabla_j\!  \times\! {\bf E}_{\gamma_2}({\bf R}_j))  + 1 \leftrightarrow 2] .
\end{eqnarray}
Note that the factor involving the summation over excited states is the ground state static polarizability of an isotropic atom, $\alpha_0=\sum_{e} (2 d_{eg}^2/3\epsilon_0 \hbar \omega_{eg}) $. 
As expected, this expression for the transition amplitude is identical to the one that results from the effective Hamiltonian approach and first-order time-dependent perturbation theory. As already mentioned, the merit of the presented approach is that it highlights atom-field dynamical processes that remain obscured in the approach that starts out from polarizabilities. 
The rate of production of photon pairs in modes $\gamma_1,\gamma_2$ is obtained taking the long time limit
$r^{(\Phi)}_{\gamma_1,\gamma_2} = \lim_{t\rightarrow \infty} (1/t)
| c^{(\Phi)}_{\gamma_1,\gamma_2}(t) |^2$, which is to be understood within time-dependent perturbation theory: $t$ is typically not longer than a fraction of the relevant atomic life times. Finally,
\begin{eqnarray}
r^{(\Phi)}_{\gamma_1,\gamma_2} =
 \frac{\pi \alpha_0^2  \omega_1 \omega_2 \Omega^2 \Delta^2}{8c^2 V^2} \delta(\omega_1 + \omega_2 - \Omega) 
\Big| \sum_j e^{- i \Phi({\bf R}_j)} W_{\gamma_1,\gamma_2}({\bf R}_j) \Big|^2.
\label{rate-final}
\end{eqnarray}
The Dirac delta ensures energy conservation in the DCE process and the last factor 
contains other conservation laws that depend on the synthetic phase, as described below.
This factor has the form of an array form factor akin to classical antenna theory:
Each atom emits with a phase $e^{- i \Phi({\bf R}_j)}$ weighted by  $W_{\gamma_1,\gamma_2}({\bf R}_j)$.
Interestingly, the atomic array is like as  driven quantum antenna emitting DCE photon pairs.

\begin{figure}[t!]
\includegraphics[width=0.8\linewidth]{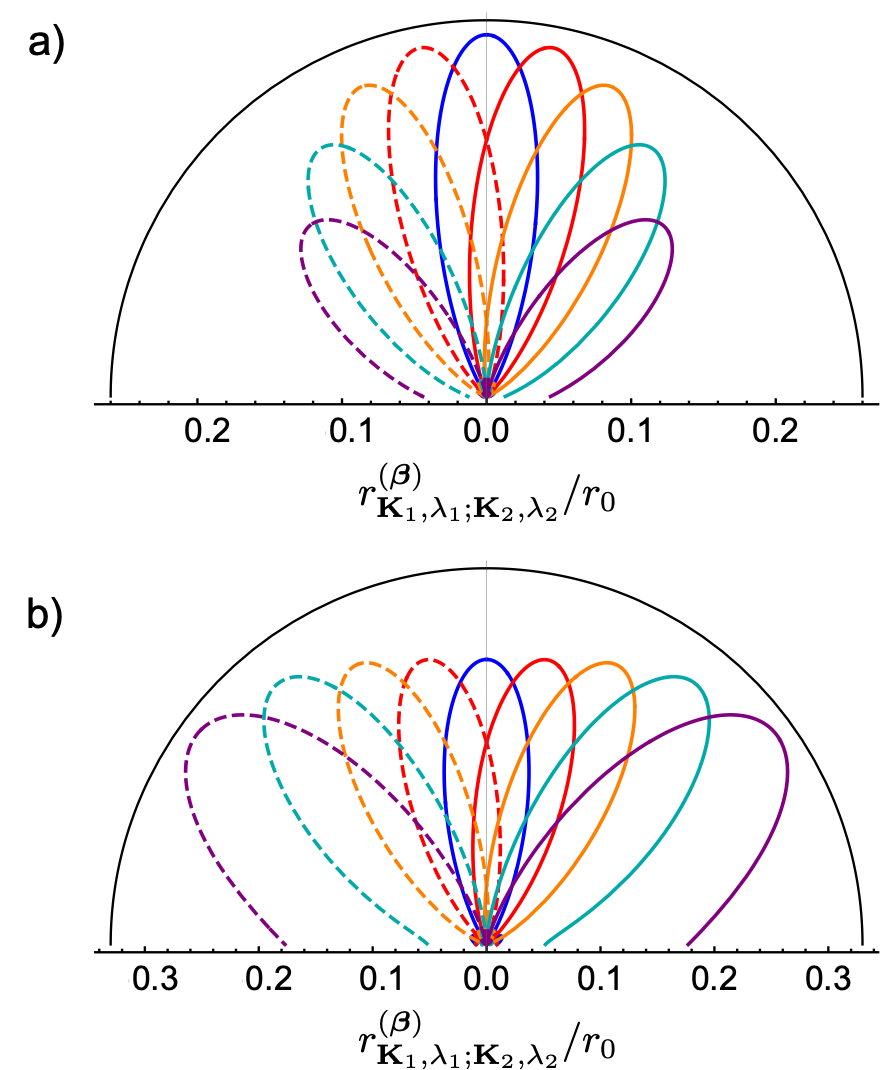}
\caption{
{\bf Two-photon angular emission.} 
Emission lobes in the $yz$-plane of one of the DCE emitted photons with (a) TE and (b) TM polarization for various magnitudes of the momentum kick $\boldsymbol{\beta} = \beta \hat{{\bf x}}$, namely 
$c \beta/\Omega = 0$ (blue), $\pm 0.1$ (red), $\pm 0.2$ (orange), $\pm 0.3$ (cyan), and $\pm 0.4$ (purple). Solid (dashed) curves correspond to positive (negative) values of $\beta$ . The other photon is assumed to be emitted along the $z$-direction and we trace over its polarization degree of freedom.
Parameters are: $N_x=N_y=50$, $N_z=1$, $ L_x=L_y=20c/\Omega$, and 
$\omega_1=\omega_2 = \Omega/2$. The emission rate is normalized to 
$r_0 = \alpha^2_0  \Omega^3 \Delta^2 / 16 (2\pi)^3 c^2 N_x^2N_y^2 N_z^2$.
}
\label{Fig2}
\end{figure}

\section{Results}

\subsection{Linear synthetic phase}

In the case 
$\Phi({\bf R})=\boldsymbol{\beta} \cdot {\bf r}$, we quantize the electromagnetic field using plane-wave 
modes labelled by $\gamma=\{{\bf K},\lambda\}$, where ${\bf K}=({\bf k},k_z)$ is the wave-vector with ${\bf k}$ its projection on the $x-y$ plane,  $k_z$ the normal projection, and $\lambda$ the polarization state. The field modes are ${\bf E}^{(\gamma)}_{\omega}({\bf R})= e^{i ({\bf k}\cdot {\bf r}+ k_z z)} \hat{\bf e}_{{\bf K},\lambda}$,
where $\hat{\bf e}_{{\bf K},\lambda}$ are the polarization unit vectors and
$\omega^2/c^2=|{\bf k}|^2+ k^2_z$ is the dispersion relation. 
In the continuum limit for the momenta $\sum_{\bf K} \rightarrow V (2\pi)^{-3} \int d{\bf K}$, the two-photon emission rate into modes ${\bf K}_1,\lambda_1;{\bf K}_2,\lambda_2$ is
\begin{eqnarray}
r^{(\boldsymbol{\beta})}_{{\bf K}_1,\lambda_1;{\bf K}_2,\lambda_2} &=&
\frac{\alpha^2_0 \omega_1 \omega_2 \Omega^2 \Delta^2 }{16 (2\pi)^5 c^2}
\delta(\omega_1\!+\!\omega_2\!-\!\Omega)
|\tilde{W}_{{\bf K}_1,\lambda_1;{\bf K}_2,\lambda_2}|^2  \cr
&& \times
\Big| \sum_j  e^{i  ({\bf k}_1+{\bf k}_2 - \boldsymbol{\beta}) \cdot {\bf r}_j } \Big|^2
\Big| \sum_j e^{i ({k}_{1z}+{k}_{2z}) z_j} \Big|^2,
\label{ratelinear}
\end{eqnarray}
where 
\begin{eqnarray}
\tilde{W}_{{\bf K}_1,\lambda_1;{\bf K}_2,\lambda_2} &=& \frac{c}{\Omega}
\Big(\frac{k_{1z} K_2}{K_1} +  \frac{k_{2z} K_1}{K_2} \Big) 
(\hat{\bf e}_{{\bf K}_1,\lambda_1} \cdot \hat{\bf e}_{{\bf K}_2,\lambda_2}) \cr
&&
- (\hat{\bf K}_2 \cdot  \hat{\bf e}_{{\bf K}_1,\lambda_1} ) (\hat{\bf z} \cdot \hat{\bf e}_{{\bf K}_2,\lambda_2}) -
(\hat{\bf K}_1  \cdot  \hat{\bf e}_{{\bf K}_2,\lambda_2} ) (\hat{\bf z} \cdot \hat{\bf e}_{{\bf K}_1,\lambda_1}) 
\nonumber \\
\end{eqnarray}
with $\hat{\bf K}_{1,2}= {\bf K}_2/K_{1,2}$ and $K_{1,2}=\omega_{1,2}/c$.
Note that $\tilde{W}_{{\bf K}_1,\lambda_1;{\bf K}_2,\lambda_2}=\tilde{W}_{{\bf K}_2,\lambda_2;{\bf K}_1,\lambda_1}$. The rate is equal to the one for a single oscillating atom (first line in Eq. (\ref{ratelinear})) multiplied by a multi-atom correction (second line) that has the form of a product of two array form factors.

In principle, each atom emits independently of its neighbors, resulting in incoherent emission of DCE photon pairs that should scale linearly in the number of atoms in the array. However, since EM quantum fluctuations have all possible wavelengths, there are large-wavelength fluctuations that coherently couple to all atoms and coherent emission \`{a} la super-radiance should be possible. In this case, the rate should scale as the square of the number of atoms, which we now show that is indeed the case. To get a close and simple expression for the form factors,  we assume atoms are arranged into a finite size cubic array with inter-atom distance $d$.  We choose the coordinate system so that the  static positions of the atoms are 
${\bf R}_j =  d m_x \hat{\bf x} +  d m_y \hat{\bf y} + d m_z \hat{\bf z}$ ($1 \le m_i \le N_i$ with $d N_i=L_i$  the size of the array in each direction). We compute the modulus square of each of the summations in Eq. (\ref{ratelinear}) and 
express the result as $AF_1({\bf k}_1,{\bf k}_2;\boldsymbol{\beta}) \; AF_2(k_{1z},k_{2z})$, where the array form factors are
\begin{eqnarray}
AF_1({\bf k}_1,{\bf k}_2; \boldsymbol{\beta}) &=&
\prod_{i=x,y} \! \frac{ \sin^2[(k_{1i}+k_{2i}-\beta_i) L_i/2]}{\sin^2[(k_{1i}+k_{2i}-\beta_i) L_i/2N_i]}, \cr
AF_2(k_{1z}, k_{2z}) &=&
\frac{ \sin^2[(k_{1z}+k_{2z}) L_z/2]}{\sin^2[(k_{1z}+k_{2z}) L_z/2N_z]} .
\label{formfactors}
\end{eqnarray}
Both array form factors are periodic functions of their arguments. 
In the $N_x, N_y \gg N_z$ limit, which mimics a finite-width large-area slab, the first array factor  is approximately equal to a two-dimensional comb of sharp peaks, all of equal height and proportional to the {\it square} of the total number of in-plane atoms, 
$(N_x N_y)^2$. This indicates that in the large-$N$ limit the emission is coherent. 
Figure 2 shows how the momentum kick controls the directionality of the emitted DCE pairs.
In the figure,  the emission direction of one of the photons is fixed to be vertical (${\bf k}_2=0$) and we trace over its polarization state. The emission lobes of the other photon are plotted for different values of $\boldsymbol{\beta}$. 
For each $\boldsymbol{\beta}$, maximal emission is when ${\bf k}_1=\boldsymbol{\beta}$.
The emission rate for TM photons is always larger than for TE photons. 

\begin{figure}[t!]
\includegraphics[width=1.0\linewidth]{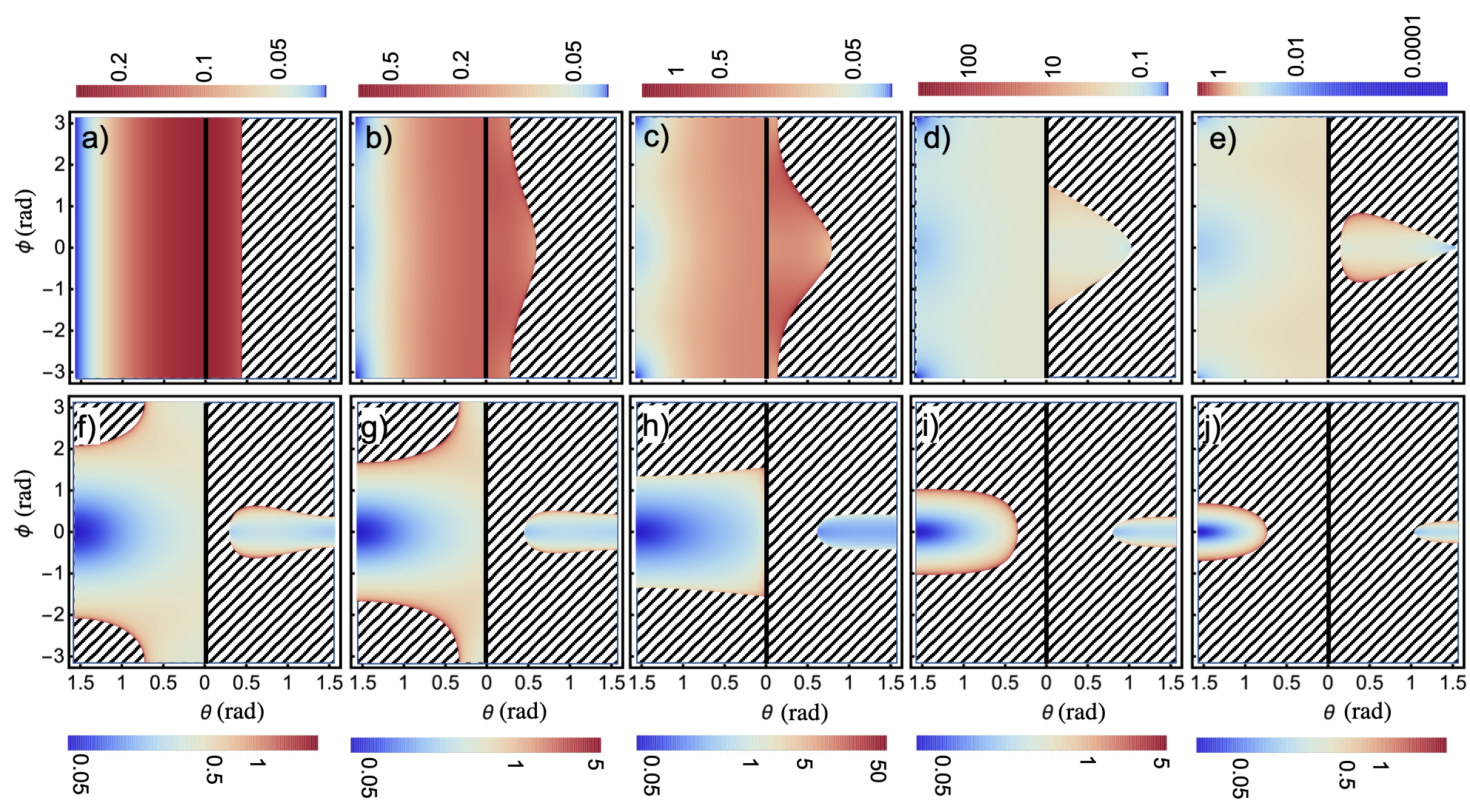}
\caption{
{\bf Density polar plots of emitted radiation.} The function $f^{(\boldsymbol{\beta})}_{\zeta,\lambda}({\bf k},\omega)$ is plotted for various momentum kicks $c\beta/\Omega$ varying from 0 to 0.9 in steps of 0.1 from (a) to (j), respectively, for the infinite periodic monolayer for TE polarization. Emission is
independent of $\zeta$, i.e., emission towards $z>0$ and $z<0$ are identical.
The case of TM emission shows similar density plots. The angles $\varphi$ and $\theta$ are the usual polar and azimuthal angles. The areas to the right (left) of the central vertical solid line correspond to the emission of the high- (low-) frequency photon in a pair with $\omega_1 = 0.7 \Omega$ and $\omega_2 = 0.3 \Omega$.  Shaded areas highlight regions of forbidden two-photon emission.
}
\label{Fig3}
\end{figure}

For an infinite atomic array in the $x-y$ plane and finite along the $z$ direction, i.e., mimicking a finite-width infinite-area slab, 
one can calculate the summation over in-plane atomic positions taking advantage of periodicity. In this case we use the lattice summation identity
\begin{equation}
\Big| \!\sum_j \! e^{i( {\bf k}_1 + {\bf k}_2- \boldsymbol{\beta}) \cdot {\bf r}_j} 
\!\Big|^2 \!\!=\!\! (2\pi)^2 A  n^2_{S} \sum_{{\bf q}}
\delta({\bf k}_1 +{\bf k}_2 - \boldsymbol{\beta}+ {\bf q}),
\end{equation}
where the sum over ${\bf q}$ corresponds to reciprocal momenta. Here, $A$ is the (infinite) area of the slab and  $n_{S}=N_x N_y/A=1/d^2$ is the number surface density of atoms.  When all relevant wavelengths are much larger than the inter-atomic distance $d$, one can take the continuous limit approximation 
in which only ${\bf q}=0$ survives the sum over ${\bf q}$. Non-zero values of the reciprocal momentum correspond to high-order diffraction modes that are evanescent and do not contribute to the generation of DCE  photons. Then we obtain 
the in-plane linear momentum conservation condition  
\begin{equation}
{\bf k}_1+ {\bf k}_2=\boldsymbol{\beta} .
\end{equation}
The time-evolved quantum state can be written as 
\begin{equation}
|\psi^{(\boldsymbol{\beta})}(t)\rangle  = 
\sum_{\bf k}  \int_0^{\Omega} \!\! d\omega  c^{(\boldsymbol{\beta})}_{\omega, {\bf k}}(t)   
|\omega, \Omega-\omega \rangle \otimes | {\bf k},\boldsymbol{\beta}-{\bf k}  \rangle,
\end{equation}
and corresponds to a color-path entangled superposition. 

In the absence of kick ($\boldsymbol{\beta}=0$), the angular spectrum of the emitted photons is always symmetric with respect to the normal of the slab, ${\bf k}_1=- {\bf k}_2$, which is what happens in the standard DCE problem of a rigidly oscillating mirror \cite{Souza2018}.
In stark contrast, when $\boldsymbol{\beta}\neq 0$, the traveling-wave modulation generates directed ripples on the mirror and produces photons that are emitted asymmetrically.
This steered DCE emission can be interpreted as a modulation-induced asymmetry of the quantum vacuum.

The DCE emission rate  is obtained by  integrating out one of the photons in the emitted pair,
$r^{(\boldsymbol{\beta})}_{{\bf K}_1,\lambda_1}=  \sum_{\lambda_2}  \int d{\bf K}_2  
r^{(\boldsymbol{\beta})}_{{\bf K}_1,\lambda_1;{\bf K}_2,\lambda_2}$.
For the finite-width infinite-area atomic array, the in-plane momentum ${\bf k}_2$ is fixed by the momentum conservation condition ${\bf k}_2 = \boldsymbol{\beta}-{\bf k}_1$, and the out-of-plane momentum is also fixed by the dispersion relation and the energy conservation condition,
$k_{2z}= \zeta_2 [(\Omega-\omega_1)^2/c^2-|\boldsymbol{\beta}-{\bf k}_1|^2]^{1/2}$, where $\zeta_2=\pm 1$ gives
the two possible emission directions normal to the array ($z>0$ or $z<0$). 
Hence, the 
momentum integration above can be done straightforwardly and only the summation over polarization states remains. Note that propagative DCE photons can be emitted only when 
$k_{1z}=\zeta_1 (\omega_1^2/c^2-|{\bf k}_1|^2)^{1/2}$ and $k_{2z}$ are real.
The spectral photon emission rate per unit area for photons with polarization $\lambda_1$, in-plane momentum in the interval $({\bf k}_1, {\bf k}_1+d{\bf k}_1)$, out-of-plane direction $\zeta_1$, and frequency in the interval $(\omega_1, \omega_1+d\omega_1)$, takes the form 
\begin{equation}
\frac{d {\Gamma}^{(\boldsymbol{\beta})}_{\zeta_1,\lambda_1}({\bf k}_1,\omega_1)}{d{\bf k}_1 d\omega_1} =
\frac{A n^2_{S} \alpha^2_0   \Delta^2 \Omega^4 \omega_1}{16 (2\pi)^3 c^5 |k_{1z}|}
f^{(\boldsymbol{\beta})}_{\zeta_1,\lambda_1}({\bf k}_1,\omega_1),
\end{equation}
where
\begin{eqnarray}
f^{(\boldsymbol{\beta})}_{\zeta_1,\lambda_1}({\bf k}_1,\omega_1)
=  \frac{\omega_1 (\Omega-\omega_1)^2}{\Omega^2 c |k_{2z}|} 
  \sum_{\zeta_2,\lambda_2} AF_2(\zeta_1 |k_{1z}|, \zeta_2 |k_{2z}|) \tilde{W}^2_{{\bf K}_1,\lambda_1;{\bf K}_2,\lambda_2}. 
\end{eqnarray}

\begin{figure}[t!]
\includegraphics[width=1.0\linewidth]{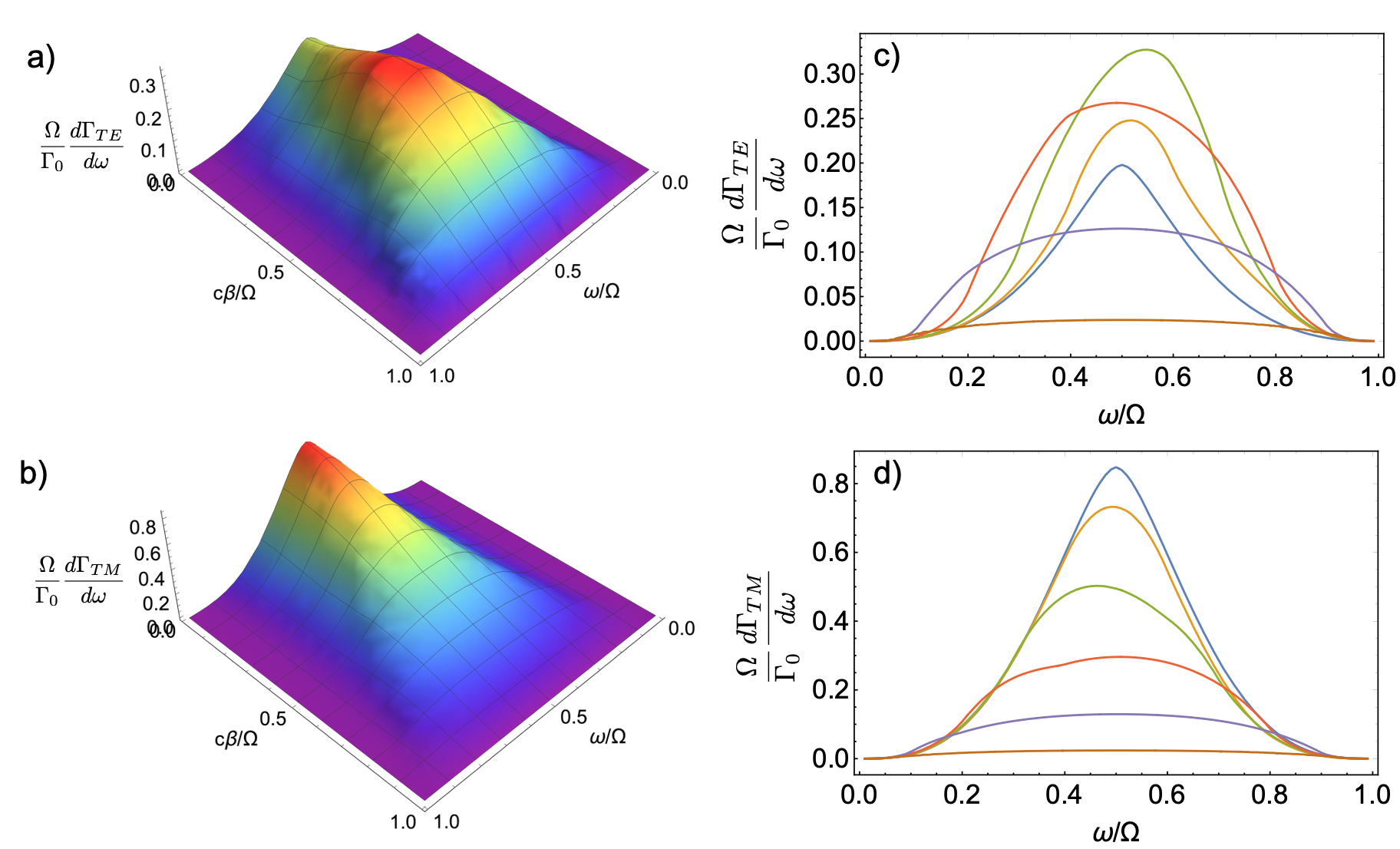}
\caption{
{\bf Tailored DCE spectrum.} The spectral emission rate of DCE photons versus frequency and momentum kick for (a) TE and (b) TM photons. Panels (c) and (d) show cuts of the spectrum for TE and TM polarizations, respectively, at fixed  $c\beta/\Omega = 0$ (blue), $0.2$ (orange), $0.4$ (green), $0.6$ (red), $0.8$ (purple), and $0.95$ (brown). We assume a monolayer with in-plane periodic boundary conditions.
The spectral rates are normalized to 
$\Gamma_0/\Omega = A n_S^2\alpha^2_0  \Omega^6 \Delta^2 / 16 (2\pi)^3 c^6$.
}
\label{Fig4}
\end{figure}

We perform the calculations in the linear polarization basis.
The emitted photons can have both $TE$ or both $TM$ polarization, and emission of photons with non-collinear in-plane momenta can be cross-polarized because
\begin{equation}
\tilde{W}^2_{{\bf K}_1,TE;{\bf K}_2,TM}  \sim [\hat{\bf z} \cdot ({\bf k}_1 \times {\bf k}_2)]^2 .
\end{equation}
In space-time DCE this can be non-zero since 
$ {\bf k}_2=  \boldsymbol{\beta}-{\bf k}_1$, while cross-polarized emission is not possible in standard DCE because photons have collinear in-plane momenta, ${\bf k}_1=-{\bf k}_2$. 
Figure 3 depicts $f^{(\boldsymbol{\beta})}_{\zeta,\lambda}({\bf k},\omega)$ for the case of an infinite monolayer,
$\lambda=TE$ polarization and $\zeta=1$ (by symmetry, the plots for $\zeta=-1$ are identical).
Polar density plots are shown both for the high-frequency ($\omega_1>\Omega/2$) and low-frequency ($\omega_2<\Omega/2$) photons in the emitted pair. In the absence of kick, the high-frequency photon can be emitted in any azimuthal direction but it has a maximum polar angle of emission (panel a, area
to the right of the central vertical solid line), while no such a constraint exists for the low-frequency photon
(panel b, area to the right of the central vertical solid line).
As the magnitude  of the momentum kick $\beta$ increases, the distributions undergo intricate changes. 
The region of allowed emission for the first photon gets deformed when the kick is non-zero and at a critical value $c \beta=\Omega-\omega_1$ (between panels d and e) an ``island" of emission appears surrounded by a sea of forbidden emission directions (shaded areas). The island drifts to higher polar angles until it touches the grazing emission line
when $c \beta = 2\omega_1-\Omega$ (between panels e and f). The island starts to shrink in size
(panels f-j), and finally at  $\beta_{max}=\Omega/c$ it collapses to a point (after panel j, not shown) and the photon is only emitted parallel to the kick. Far-field emission above that value of the kick is not possible. Regarding the second photon, its emission distribution remains mostly unperturbed until 
at $c \beta=\Omega-2 \omega_2$ two areas of forbidden emission appear at large polar angles and opposite to the kick direction (between panels e and f). The forbidden region grows until it engulfs its allowed emission region and a second island forms at $c \beta=\Omega-\omega_2$ (between panels h and i). Finally, it ends up being emitted at a grazing angle but in a direction anti-parallel to the kick (after panel j, not shown). The modulation also excites  hybrid entangled pairs composed of one  photon and one evanescent surface wave (shaded areas), and when $\beta>\beta_{max}$ only evanescent modes are created  on the atomic monolayer and subsequently decay via non-radiative loss mechanisms. 

\begin{figure}[t!]
\includegraphics[width=1.0\linewidth]{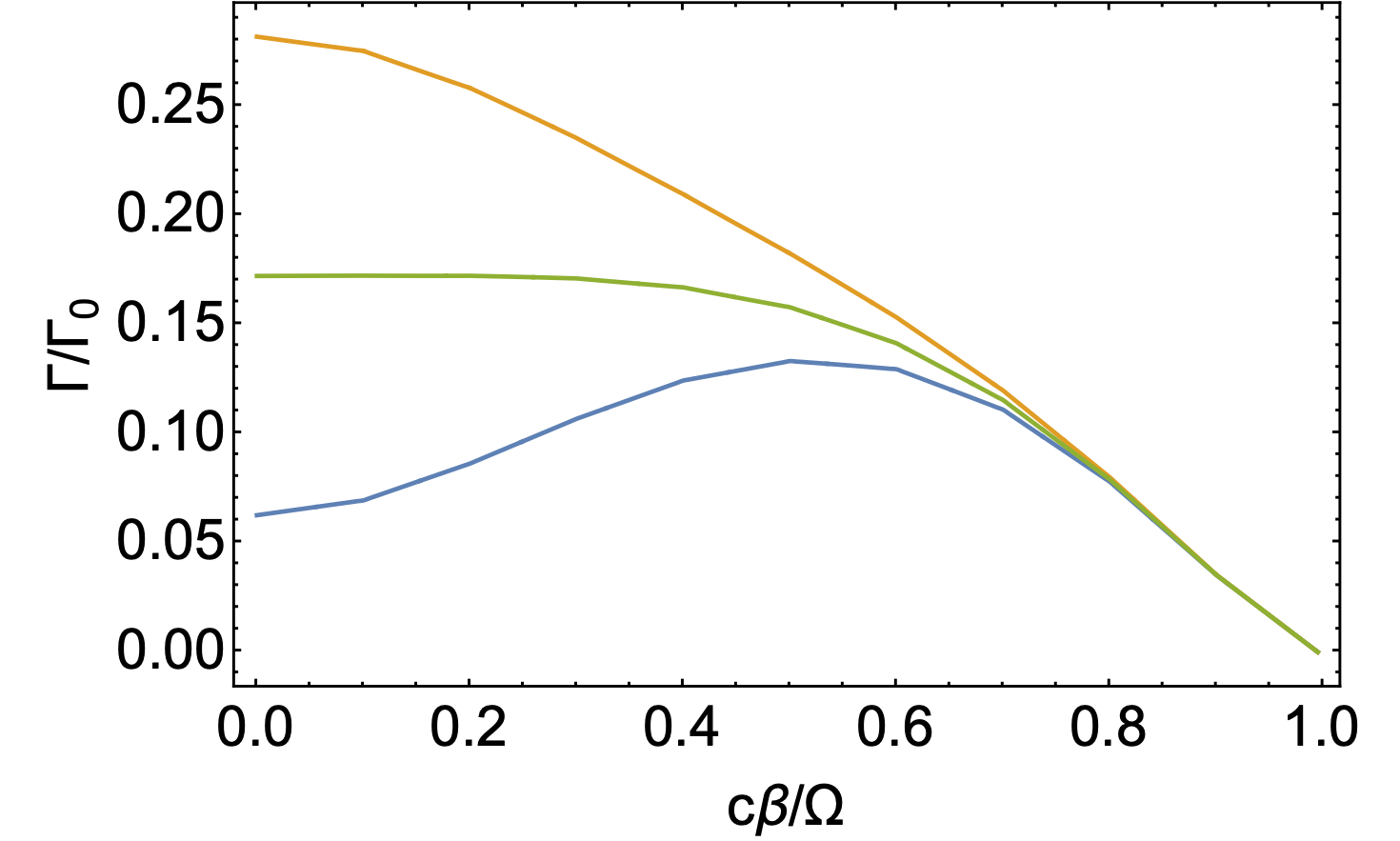}
\caption{
{\bf DCE emission rate}. Total photon creation rate for the system considered in Fig. 4 for 
TE (blue), TM (orange), and circularly (green) polarized photons.}
\label{Fig5}
\end{figure}

The DCE spectral emission rates for photons with polarization $\lambda$ are obtained by 
integrating over all emission directions
\begin{equation}
\frac{d\Gamma_{\lambda}(\omega,\beta)}{d\omega} = \frac{A n^2_{S} \alpha^2_0  \Omega^4 \Delta^2 }{16 (2\pi)^3 c^5 } \sum_{\zeta} \int d{\bf k} \frac{  \omega f^{(\boldsymbol{\beta})}_{\zeta,\lambda}({\bf k},\omega)}{|k_{1z}|}
\end{equation}
and depends only on the modulus of the kick.
The TE and TM spectral emission rates are shown in Fig. 4 as a function of frequency and momentum kick.
At zero momentum kick, they are both symmetric with respect to $\omega=\Omega/2$ and the rate of 
TE emission is smaller than TM emission (note the different vertical scales in the plots).
The two rates are identical to those derived in \cite{Souza2018} for the standard DCE problem in the absence of synthetic phase. At non-zero kick, they both become slightly asymmetric,  with the peak of the TE (TM) moving to frequencies larger (smaller) than $\omega=\Omega/2$. 
The origin of the asymmetry is the non-zero cross-polarized emission.
Indeed, the TE spectral emission rate has contributions from two terms, 
$\tilde{W}^2_{{\bf K}_1,TE;{\bf K}_2,TE}$ and $\tilde{W}^2_{{\bf K}_1,TE;{\bf K}_2,TM}$.
The first term is symmetric around $\omega=\Omega/2$ because upon interchanging the frequency, momentum, and spin indices one gets
$\tilde{W}^2_{{\bf K}_1,TE;{\bf K}_2,TE}=\tilde{W}^2_{{\bf K}_2,TE;{\bf K}_1,TE}$. However, 
the second term is not symmetric because the same interchange gives
$\tilde{W}^2_{{\bf K}_1,TE;{\bf K}_2,TM} = \tilde{W}^2_{{\bf K}_2,TM;{\bf K}_2,TE}
\neq \tilde{W}^2_{{\bf K}_2,TE;{\bf K}_2,TM}$.  The same argument applies to the TM spectral emission rate.
The spectral emission rate summed over polarizations 
is symmetric, though.

The total rate per polarization is obtained by integrating over frequency
\begin{equation}
\Gamma_{\lambda}(\beta) = \int_0^{\Omega} d\omega  \frac{d\Gamma_{\lambda}(\omega,\beta)}{d\omega}.
\end{equation}
As shown in Fig. 5, the TM rate shows a monotonous decay to zero  at $\beta_{\max}=\Omega/c$, while
the TE rate decays non-monotonically. Furthermore, $\Gamma_{TE}(\beta)<\Gamma_{TM}(\beta)$ and their difference diminishes as the kick grows.  
The figure also shows the emission rate for circularly polarized photons, 
which is the same for right- and left-circular polarization, 
$\Gamma_{R}(\beta) =\Gamma_{L}(\beta)$, and is initially flat and then decays monotonically to zero.
As follows from the figure, 
$\Gamma_{TE}(\beta)<\Gamma_{R/L}(\beta)<\Gamma_{TM}(\beta)$  and the total rate
verifies $\Gamma(\beta)=\Gamma_{TE}(\beta)+\Gamma_{TM}(\beta)=2\Gamma_{R/L}(\beta)$.


\subsection{Spinning synthetic phase}

In this section we briefly discuss the case of a spinning synthetic phase $\Phi({\bf R})=\ell \varphi$. 
Because of the symmetry properties of the phase, one needs to quantize the electromagnetic field with angular momentum.
Usually, this is done in the paraxial approximation in terms of Laguerre-Gauss modes that carry orbital angular momentum \cite{Calvo2006}. However, for DCE this is not appropriate because the emitted photon pairs are non-paraxial, and a more general approach is required. We follow the quantization scheme based on vector-Bessel modes, that form a complete basis for the electromagnetic field with angular momentum and do not require any paraxial approximation \cite{Enk1994}. In this scheme, neither orbital angular momentum (OAM) nor spin angular momentum (SAM) are good quantum numbers, only their sum is. 
Vector-Bessel  modes are labelled by $\gamma=\{ k, k_z,\eta,m \}$, where $k\ge 0$ is the transverse linear momentum,
$k_z$ is the axial linear momentum, $\eta=\pm 1$ is associated with the sign of the transverse spin $s_t= \pm \hbar ck_z/\omega$, and
$m$ is the total angular momentum (OAM plus SAM).
The dispersion relation $\omega^2/c^2=k^2+k^2_z=K^2$ does not depend on $m$ or $\eta$.
In cylindrical coordinates, the vector-Bessel mode is 
${\bf E}^{(\gamma)}_{\omega} = E_{\rho} \hat{\boldsymbol{\rho}} + E_{\varphi } \hat{\boldsymbol{\varphi}}
+ E_z \hat{\bf z}$, where
\begin{eqnarray}
E_{\rho}\! &=&\! \frac{i e^{i k_z z} e^{i m \varphi}}{\sqrt{2} (2\pi) } 
\!\left[ \!
\frac{k_z \! +\!  \eta K}{2 K} J_{m-1}(k \rho) \!-\!
\frac{k_z \! -\!  \eta K}{2 K} J_{m+1}(k \rho) 
 \right]  \cr
E_{\varphi} \! &=&\! 
-\frac{e^{i k_z z} e^{i m \varphi}}{\sqrt{2} (2\pi)} 
\!\left[ \!
\frac{k_z \! +\!  \eta K}{2 K} J_{m-1}(k \rho) \!+\!
\frac{k_z \! -\! \eta K}{2 K} J_{m+1}(k \rho) 
\!\right] \cr
E_{z} \! &=&\!  \frac{e^{i k_z z} e^{i m \varphi}}{\sqrt{2}(2\pi)}  \frac{k}{K} J_m(k \rho),
\end{eqnarray}
where  $J_{m}(x)$ are Bessel functions. Note these modes are non-diffracting (constant amplitude along the $z$ direction), have a topological vortex singularity along the $z$-axis with a phase wrapping equal to $2 \pi m$, and decay along the radial direction. 

The two-photon generation rate is 
\begin{eqnarray} r^{(\ell)}_{\gamma_1, \gamma_2}\! =\!
\frac{\pi \alpha^2_0 \omega_1 \omega_2 \Omega^2 \Delta^2 }{8 (2\pi)^2 c^2}
\delta(\omega_1+\omega_2\!-\!\Omega)  
\label{ratespinning} 
\Big|\! \sum_j e^{i ({k}_{1z}+{k}_{2z}) z_j}\Big|^2
\Big|\! \sum_j T_{\gamma_1, \gamma_2}(\rho_j)  e^{i (m_1+m_2-\ell) \varphi_j}\Big|^2,
\end{eqnarray}
where we took the continuum limit $\sum_{k,k_z} \rightarrow V \int_{-\infty}^{\infty}  d k_z  \int_0^{\infty} k dk$. The function $T_{\gamma_1, \gamma_2}(\rho_j)$ results from plugging the vector Bessel mode into Eq. (\ref{rate-final}) and  has  a cumbersome expression that we do not write here (it will be written below after performing the summation over $j$). Emitted photon pairs can have the same ($\eta_1 \eta_2=1$) or opposite
($\eta_1 \eta_2=-1$) signs of transverse spin.
In contrast to the rate for the linear synthetic phase Eq. (\ref{ratelinear}), it is not possible to express the rate 
Eq. (\ref{ratespinning})
as that of a single atom multiplied by a multi-atom correction.  The underlying reason is the nontrivial topology of the imprinted modulation, which is ill-defined for a single atom. 

To perform the sum over atoms, we assume they are arranged into a cylindrical geometry of radius $R$ and height $L_z$, forming
parallel disks along the $z$ direction (inter-disk distance $d$) and concentric rings on the orthogonal plane
(inter-ring separation $r$). The axis of the cylinder coincides with the axis of spinning modulation axis ($z$-direction).  The second line in Eq. (\ref{ratespinning}) is the product of two array form factors:
one is the same  $AF_2$ discussed in the previous section, and the other is a new
$AF_3$ that depends on all quantum numbers $k_1,k_2; k_{1z}, k_{2z}; \eta_1,\eta_2; m_1, m_2$ of the two photons. We compute  $AF_3$ by writing 
the radial coordinates of the atoms as $\rho_j=r n$, with  $0 \le n \le N_R$  and $r N_R = R$, and the azimuthal coordinates on the $n-$th ring as $\varphi_j(n)= (2 \pi/N(n)) p$ with $p=0, 1, ..., N(n)-1$. We assume the number of atoms on a given ring is proportional to the ring perimeter, hence $N(n)= \xi n$ with $\xi$ a constant. In the limit $\xi \gg 1$, the atoms form a quasi-continuum on the azimuthal direction and the summation over the azimuthal index $p$ gives 
$| \sum_{n,p} T(\rho_j) e^{i(m_1+m_2-\ell) \varphi_j}|^2 = \delta_{m_1+m_2-\ell} | \sum_n N(n) T(r n)|^2$.
Hence, the spinning modulation stirs the quantum vacuum and generates vortex photon pairs whose angular momentum must add up to the imprinted topological charge $\ell$:
\begin{equation}
m_1+m_2=\ell
\end{equation}
in agreement with angular momentum conservation. For $\ell=0$ the generated photons in a pair have opposite twist. The two-photon state is frequency-angular momentum entangled,
\begin{equation}
|\psi^{(\ell)}(t)\rangle  = 
\sum_{m}  \int_0^{\Omega} \!\! d\omega \;  c^{(\ell)}_{\omega,m}(t)   
|\omega, \Omega-\omega \rangle \otimes |m,\ell-m  \rangle.
\end{equation}
The remaining summation over the radial coordinates
is performed in the limit of a continuum distribution of atoms, $r \rightarrow 0$ and $N_R \rightarrow \infty$ such that $r N_R=R$ is constant. We define $x=n r$ and approximate the sum 
as an integral over the continuous variable $x$. The array form factor takes the final form
\begin{eqnarray}
&& AF_{3} =  (2\pi)^2 n_S^2   \delta_{m_1+m_2, \ell} \; \Big[\int_0^R \!\!dx \; x T_{\gamma_1, \gamma_2}(x) \Big]^2,
\end{eqnarray} 
where $n_S$ is the number of atoms per unit of disk area.
The total emission rate  from the modulated atomic array  is
\begin{equation}
\Gamma_{\ell} =\frac{\pi \alpha^2_0  \Omega^2 \Delta^2}{8 (2\pi)^2 c^6} \int_0^{\Omega} d\omega  \; \omega^2 (\Omega-\omega)^2 f_{\ell}(\omega) .
\label{rate-spinning}
\end{equation}
The spectral weight function $f_{\ell}(\omega)$ sums the product
$AF_2 AF_3$ over all degrees of freedom of the two photons except the frequency of one of them.
In the case of zero imprinted angular momentum, the total rate is identical to that derived in \cite{Souza2018} for the standard DCE problem: $\Gamma_{\ell=0} = \Gamma_{\boldsymbol{\beta}=0} =\Gamma_{\Phi=0}$.

To analyze the angular momentum content of the emitted radiation, we expand
the spectral weight function into different $m$ contributions, 
$f_{\ell}(\omega)=  [ \pi^2 n_S^2 c^4/\omega^2 (\Omega-\omega)^2] \sum_{m} f_{\ell}(\omega,m)$.
The angular momentum spectrum is 
\begin{equation}
f_{\ell}(\omega,m) =
\int_0^{u}  d\kappa \int_0^{1-u}  d\kappa' 
\sum_{\zeta,\zeta'=\pm1} \;
\sum_{\eta,\eta'=\pm1} ( {t}_a +{t}_b+{t}_c+{t}_d+t_e)^2,
\end{equation}
where the different terms are written in dimensionless variables
$u=\omega/\Omega$,
$\kappa_z=\zeta [ u^2 -  \kappa^2]^{1/2}$, 
$\kappa'_z=\zeta' [ (1-u)^2 -  (\kappa')^2]^{1/2}$
and 
${\cal R}=\Omega R/c$, as
\begin{eqnarray}
{t}_a&= &
\Big[ 
[ \kappa_{z}  + \kappa'_{z} (1+(1-u)^{-1}) ] 
(\kappa_{z} + \eta u) (\kappa'_{z} -\eta' (1-u))\cr 
&+& 
\frac{(\kappa')^2 (\kappa_{z} +  \eta u) }{2 (1-u)} + \frac{\kappa^2  (\kappa'_{z} - \eta' (1-u))}{2u} 
\Big] {\cal H}^{(-)}_{m,\ell} ,
\end{eqnarray}
\begin{eqnarray}
{t}_b&= &
\Big[ 
[ \kappa'_{z}  + \kappa_{z} (1+u^{-1}) ] 
(\kappa'_{z} + \eta' (1-u)) (\kappa_{z} -\eta u)\cr
 &+& 
\frac{\kappa^2 (\kappa'_{z} +  \eta' (1-u)) }{2 u} + \frac{(\kappa')^2  (\kappa_{z} - \eta u)}{2 (1-u)} 
\Big] {\cal H}^{(+)}_{m,\ell} ,
\end{eqnarray}
\begin{eqnarray}
{t}_c &= &
\Big[ \kappa_{z}  (2+u^{-1})+ \kappa'_{z} (2+ (1-u)^{-1}) 
 \Big] \kappa \kappa'  {\cal H}^{(0)}_{m,\ell} ,
 \end{eqnarray}
\begin{eqnarray}
{t}_d &=& 
- \Big[ \frac{(\kappa')^2 (\kappa_{z} +  \eta u) }{2 (1-u)} + \frac{\kappa^2  (\kappa'_{z} - \eta' (1-u))}{2u} \Big]
{\cal I}^{(-)}_{m,\ell}\cr
 &-&
\Big[ \frac{(\kappa')^2 (\kappa_{z} -  \eta u) }{2 (1-u)} + \frac{\kappa^2  (\kappa'_{z} + \eta' (1-u))}{2u} \Big]
{\cal I}^{(+)}_{m,\ell},
\end{eqnarray}
\begin{eqnarray}
{t}_e &=& 
\frac{m \kappa (\kappa'_{z} +  \eta' (1-u))  }{u}   {\cal J}^{(-)}_{m,\ell} + 
\frac{m \kappa(\kappa'_{z} - \eta' (1-u) ) }{u}   {\cal J}^{(+)}_{m,\ell} \cr
& +&
\frac{(\ell-m) \kappa'  (\kappa_{z} +  \eta u) }{1-u} {\cal K}^{(-)}_{m,\ell} + 
\frac{(\ell-m) \kappa'(\kappa_{z} - \eta u)}{1-u}   {\cal K}^{(+)}_{m,\ell} .
\end{eqnarray}
Here, we defined 
dimensionless integrals that are functions of $\kappa$, $\kappa'$ and ${\cal R}$:
\begin{eqnarray}
{\cal H}^{(\pm)}_{m,\ell} &=& - (-1)^{m-\ell} \int_0^{\cal R} \!\!\!\!dy \; y J_{m \pm 1}(\kappa y) J_{m \pm 1- \ell}( \kappa' y) \cr
{\cal H}^{(0)}_{m,\ell}&=&  (-1)^{m-\ell}  \int_0^{\cal R}   \!\!\!\!dy \; yJ_{m}( \kappa y) J_{m-\ell}( \kappa' y) \cr
{\cal I}^{(\pm)}_{m,\ell}&=& -  (-1)^{m-\ell} \int_0^{\cal R} \!\!\!\!dy \; yJ_{m \pm 1}( \kappa y) 
J_{m \mp 1 - \ell}( \kappa' y) \cr
{\cal J}^{(\pm)}_{m,\ell} &=&-  (-1)^{m-\ell} \int_0^{\cal R} \!\!\!\!dy \; J_{m}( \kappa y) J_{m\mp 1-\ell}(\kappa' y)  \cr
{\cal K}^{(\pm)}_{m,\ell} &=& (-1)^{m-\ell} \int_0^{\cal R}   \!\!\!\!dy \;  J_{m \pm1}(\kappa y) J_{m-\ell}(\kappa' y) .
\end{eqnarray}
For $\ell=0$, the integrals ${\cal H}^{(\pm)}$ and ${\cal H}^{(0)}$ are one of Lommel's integrals and have a closed form, but the other three integrals do not.
In the limit ${\cal R} \rightarrow \infty$, all the integrals are of the Weber-Schafheitlin form, $\int_0^{\infty} dy y^q J_{\mu}(\kappa y) J_{\nu}(\kappa' y)$, with exponent $q=1$ or $q=0$.  
The above expressions allow to calculate $f_{\ell}(\omega,m)$ in space-time motion-induced DCE under spinning modulation. 
Direct inspection of $f_{\ell}(\omega,m)$ shows that it satisfies the condition $f_{\ell}(\omega,m)= f_{\ell}(\Omega-\omega,\ell-m)$. This identity states the simple fact that, as photons are emitted in pairs satisfying energy and angular momentum conservation, 
the emission rate of a photon with given frequency and angular momentum must be the same as the emission rate at the complementary frequency and angular momentum.  
Also, one intuitively expects that one of the emitted photons should most probably be  emitted with an
angular momentum equal to the driving, i.e., $m=\ell$.  Because of angular momentum conservation, its companion in the pair is most probably emitted with $m=0$. This intuition was confirmed in \cite{STQM} for the case of DCE photons emitted from a quantum metasurface with spinning modulation of its optical properties.

Finally, we briefly discuss the structure of the energy density and
Poynting vector in space-time DCE with spinning synthetic phase. Their expectation values on the 
evolved quantum state  have a single vortex singularity along the $z$-axis for $\ell \neq 0$.
The reason why there is a single vortex line and not two (one per emitted photon) is due to the nature of an expectation value: it corresponds to summing over infinite many realizations of  DCE two-photon emission events, and as such it is not surprising that individual dual vortex events average out and lead to a single vortex aligned with the modulation axis. Since there is only a single preferred axis in the problem, symmetry considerations explain why dual vortices cannot occur. A more formal justification is the non-diffracting nature of the vector-Bessel modes employed in the quantization scheme. However,  a single experimental realization will have dual vortices along random directions of emission. 

\section{Discussion}

As mentioned in the introduction, motional DCE  involving temporal modulation of a boundary has not been observed to date because experimentally feasible mechanical modulation frequencies are insufficient for generating a large amount of photons per unit of time. For example,  
the rate of photon creation from an oscillating perfectly-reflecting mirror of area $A$ is
$\Gamma_{DCE} = A \Omega^5 \Delta^2 / 15 (2\pi)^2 c^4$ \cite{MaiaNeto1994}, which gives  $\sim 10^{-21}$ photons s$^{-1}$ for a mirror of 1 cm$^2$ area, modulation frequency $\Omega/2\pi \sim$ 1 MHz, and modulation amplitude $\Delta \sim 100$ nm. This is less than one photon every 10 years.
An alternative approach to moving the whole mirror is to make its surface oscillate by launching acoustic waves, a scheme that is related to the proposed space-time motional DCE. However, usual materials can bear maximal  relative deformations 
$\delta_{max} = \Delta_{max} / (v_s/\omega_w) \sim 10^{-2} $ ($v_s$ is the speed of sound and $\omega_w$ is the frequency of the acoustic wave) and have maximal velocities of the boundary 
$v_{max} \sim \delta_{max}  v_s \sim 50$ m/s, which again leads to negligibly small photo-production rates \cite{Dodonov2010}. The same limitation occurs in space-time motional DCE. According to Fig. 5, the maximal rate happens for zero synthetic phase and is 
\begin{equation}
\Gamma_{max} \approx 0.34 \Gamma_0  = \frac{0.34}{16 (2\pi)^3 }
\frac{A n_S^2\alpha^2_0  \Omega^7 \Delta^2}{c^6}.
\end{equation} 
We estimate it for experimentally feasible parameters of a spatio-temporally modulated meta-mirror
consisting of an array of $^{87}$Rb atoms (ground state polarizability $\alpha_0=5.9 \times  10^{-28} {\rm m}^3$ \cite{Steck2019}) loaded into a 2D square  optical lattice and vertically trapped by another shaking optical lattice. For a lattice constant $d= 532$nm, unit filling, $N \sim 200$ atoms ($A \sim 50 \mu {\rm m}^2$, $n_S \sim 4 \mu {\rm m}^{-2}$) \cite{Rui2020}, modulation amplitude $\Delta \sim 100$ nm, 
and modulation frequency $\Omega/2 \pi \sim 10$ kHz \cite{Lignier2007,Parker2013}, 
the maximal rate is $\Gamma_{max} \sim 10^{-77}$ photons  s$^{-1}$. Again, unrealistic large modulation frequencies would be required to get measurable DCE rates.

In order to face these severe limitations of both standard and space-time motional DCE,
analog DCE set-ups based on modulation of optical properties are required. For standard DCE, the experiment \cite{Wilson2011} based on a superconducting microwave coplanar waveguide is an analog of a one-dimensional DCE mirror, with the rate given by $\Gamma=(\Omega/12\pi){(v_{eff}/c)}^2$, where $v_{eff}$ is the effective velocity of the mirror. For the experimental parameters $\Omega/2\pi=11$ GHz and $v_{eff}/c=0.05$,
the estimated rate is
$\Gamma \sim 10^{6}$ photons  s$^{-1}$. For space-time DCE, a possible analog of a DCE mirror moving in three dimensions is a set-up consisting of a graphene-disk metasurface whose  electro-optical properties are spatio-temporally modulated.
All-optical modulation of the Fermi energy at THz frequencies should enable giant rates $\Gamma \sim 10^{12}$ photons  s$^{-1}$ from centimeter-sized metasurfaces  \cite{STQM}.

\section{Conclusions}

The space-time dynamical Casimir effect offers a novel degree of control over photo-production from the quantum vacuum.  A space-dependent synthetic phase distribution imprinted on the temporal modulation protocol 
of optical  properties or geometrical boundaries enables the generation of photon pairs with tailored spatial modes and entanglement properties. 
Travelling-wave modulations, like parallel ripples propagating on the surface of a mirror, generate steered photon pairs that are frequency-path entangled. Spinning-wave modulations, like twisting ripples on the mirror, produce vortex photon pairs featuring frequency-angular momentum entanglement. The synthetic phase can be reconfigured on-demand by changing the modulation protocol, allowing to modify the nature of entanglement between the generated photon pairs. 

We end the paper with a conjecture. As discussed above, the synthetic phase distribution $\Phi({\bf r})$ is imprinted onto space-time quantum metasurfaces via temporal delay of the modulation signal on different meta-atoms  \cite{STQM}. The meta-atoms can all have the same geometry, as the case of the atomic array meta-mirror, or the surface can even have no meta-atoms (unstructured), as is the case of a flat mirror. In both scenarios the spatial profile of the emitted DCE photons is controlled by the synthetic phase imprinted by the modulation. On the other hand, a phase distribution $\Psi({\bf r})$ can be imprinted onto a static surface by decorating it with meta-atoms having judiciously designed geometrical parameters, so-called gradient metasurfaces. For example, a blazed grating mirror or a meta-mirror with resonators of varying size. The phase distribution controls the spatial profile of light reflected from the static metasurface. An interesting question is: what is the nature of DCE emission when such gradient metasurfaces are set in motion along a direction normal to the metasurface plane?  We conjecture that when
$\Psi({\bf r})=\Phi({\bf r})$, the emitted Casimir light from the moving gradient metasurface has the same properties as the Casimir light generated by the space-time modulation on an unstructured surface.  For example, the blazed grating mirror moving orthogonal to the grating plane should produce steered frequency-path entangled photon pairs, just as a flat mirror under a spatio-temporal modulation with a linear synthetic phase would. A test of this conjecture will
require to calculate DCE emission from moving surfaces decorated with complex nanostructures, 
an overwhelmingly challenging analytical calculation and a very hard numerical task on its own. Were the conjecture proven to be correct, it would allow much simpler evaluation of DCE emission from moving structured bodies by simply considering unstructured bodies modulated with the appropriate synthetic phase. 

\vspace{6pt} 



\authorcontributions{All authors equally contributed and have read and agreed to the published version of the manuscript.}

\funding{This work was supported by the DARPA QUEST and LANL LDRD programs.}


%

\dataavailability{The data supporting this research is available from the corresponding author upon reasonable request.} 

\acknowledgments{We are grateful to A. Azad, E. Efimov, M. Julian, C. Lewis, and M. Lucero for discussions.}

\conflictsofinterest{The authors declare no conflict of interest.} 



\abbreviations{Abbreviations}{
The following abbreviations are used in this manuscript:\\

\noindent 
\begin{tabular}{@{}ll}
DCE & Dynamical Casimir Effect
\end{tabular}}

%
%

\end{paracol}
\reftitle{References}

\end{document}